\documentclass[reprint,
superscriptaddress,
amsmath,amssymb,
aps,prl]{revtex4-2}
\usepackage{braket}
\usepackage{svg}
\usepackage{amsmath}
\usepackage{hyperref}
\usepackage{dcolumn}
\newcommand{\figref}[1]{Fig.\,\ref{#1}}
\graphicspath{{Figures/}{./}}

\begin{document}

\title{Experimental realization of a fermionic spin-momentum lattice}
\author{Paul Lauria}
\affiliation{Department of Physics and Astronomy, University of California at San Diego, La Jolla, CA 92093, USA}
\author{Wei-Ting Kuo}
\affiliation{Department of Physics and Astronomy, University of California at San Diego, La Jolla, CA 92093, USA}
\author{Nigel R. Cooper}
\affiliation{T.C.M. Group, Cavendish Laboratory, University of Cambridge, Cambridge, CB3 0HE, United Kingdom}
\author{Julio T. Barreiro}
\affiliation{Department of Physics and Astronomy, University of California at San Diego, La Jolla, CA 92093, USA}
%\date{\today}

\begin{abstract}

We experimentally realize a spin-momentum lattice with a homogeneously trapped Fermi gas. The lattice is created via cyclically-rotated atom-laser couplings between three bare atomic spin states, and are such that they form a triangular lattice in a synthetic spin-momentum space. We demonstrate the lattice and explore its dynamics with spin- and momentum-resolved absorption imaging. This platform will provide new opportunities for synthetic spin systems and the engineering of topological bands. In particular, the use of three spin states in two spatial dimensions would allow the simulation of synthetic magnetic fields of high spatial uniformity, which would lead to ultra-narrow Chern bands that support robust fractional quantum Hall states.

\end{abstract}

\maketitle

Ultracold atoms in optical lattices have been established as an important tool for the quantum emulation of condensed matter models~\cite{Schaefer2020}, especially those with topological features~\cite{Goldman2016,Cooper2019}. The inherent tunability afforded by optical lattices provides access to a variety of parameter regimes, which has proved essential in the seminal realizations of topological phases in ultracold matter ~\cite{Miyake2013,Aidelsburger2013,Aidelsburger2014,Jotzu2014}. Since then, efforts to study topology in other systems have led to the exploration of synthetic dimensions~\cite{Boada2012,Celi2014}, which provide internal degrees of freedom beyond those afforded by the trapping geometry and have enabled a new generation of experiments~\cite{Ozawa2019}. 

Several approaches to synthetic dimensions have been experimentally realized. Real-space lattices augmented with spin-orbit coupling (SOC) connect spin ``lattice" sites via momentum exchange, creating Hall cylinders pierced by magnetic flux in a synthetic position-spin space~\cite{Mancini2015,Stuhl2015,Kang2018,Song2018,Genkina2019,Chalopin2020}, or creating Hall ribbons in optical clock experiments~\cite{Kolkowitz2016,Livi2016,Bromley2018,Lu2021}. Real-space lattices are not always needed; SOC itself can provide  synthetic degrees of freedom, which can act as a potent generator of Berry curvature~\cite{Meng2016,Yi2019,ValdesCuriel2021,Liang2021,Fabre2021} or provide control parameters for Hamiltonian engineering~\cite{Wang2021,Ren2021}. Synthetic lattices entirely in momentum-space~\cite{Meier2016a,An2021} have been realized, and, with carefully engineered hopping schemes, have proven topological~\cite{Meier2016,An2017,Xie2019,Xie2020}. Recently, a synthetic lattice of Rydberg states has been employed for the study of a Su–Schrieffer–Heeger model~\cite{Kanungo2021}, and a synthetic dimension of trap states created with patterned light \cite{Oliver2021}.  

Lattices composed of spin and momentum states, or spin-momentum (SM) lattices, have been proposed~\cite{Cooper2012} as a platform to exhibit topological features, with some schemes potentially realizing the Laughlin state of the fractional quantum Hall effect~\cite{Stormer1999,Cooper2013}. As a step towards this, we realize a fermionic spin-momentum lattice using SOC and three atomic Zeeman spin states. Previous experiments using spin-momentum lattices utilized bosons in a single-dimension~\cite{Anderson2020} or used a real-space lattice with lattice band pseudospins~\cite{Khamehchi2016}. Here, by providing sufficient links between spin sites, we build a lattice in a 2D spin-momentum space without a traditional scalar optical lattice. This platform increases the flexibility of the synthetic dimension approach. In particular, the use of three spin states in two spatial dimensions allows the simulation of synthetic magnetic fields of high spatial uniformity, which lead to ultra-narrow Chern bands that support robust fractional quantum Hall states~\cite{Cooper2011a,Cooper2013}. 

\emph{Implementation.} The synthetic lattice is composed of three Zeeman spin states in the ${}^1S_0(F=9/2)$ ground state of ${}^{87}\rm{Sr}$, labeled $ X \equiv |m_F=-9/2\rangle, Y \equiv |m_F=-7/2 \rangle, Z \equiv |m_F=-5/2 \rangle$. In a momentum-dependent manner, the spins are cyclically coupled by up to 9 Raman lasers intersecting at 120$^\circ$. In the rotating-wave approximation, we describe the atom-laser coupling as
\begin{equation}
\hat{V}= \Omega_{mn}e^{i(\mathbf{k}_R\cdot\mathbf{r}+\varphi_i-\varphi_j)} \lvert m \rangle\langle n \rvert +\mathrm{H.c.}
\end{equation}
where $m\neq n$ runs over the states $X,Y,Z$, $|\mathbf{k}_R|=|\mathbf{k}_i-\mathbf{k}_j| = k_L \sin^2{\frac{\theta}{2}}$ is the magnitude of the single-photon recoil wavevector with $i\neq j$ denoting the beams driving a particular $m$-$n$ coupling, $\Omega_{mn}$ is the coupling strength, and $\theta=120^\circ$ is the angle between any pair of beams. The single-photon recoil energy is $E_R= (\hbar k_R)^2/2m= \hbar \times 22.7 ~\mathrm{kHz}$. The phase differences $\varphi_i-\varphi_j$ are set to zero in the experiment, but we note that setting nonzero phases is at the heart of the ultra-narrow-band optical flux lattice experiment\cite{Cooper2013}. 

\begin{figure}[t]
\includegraphics[width=\columnwidth]{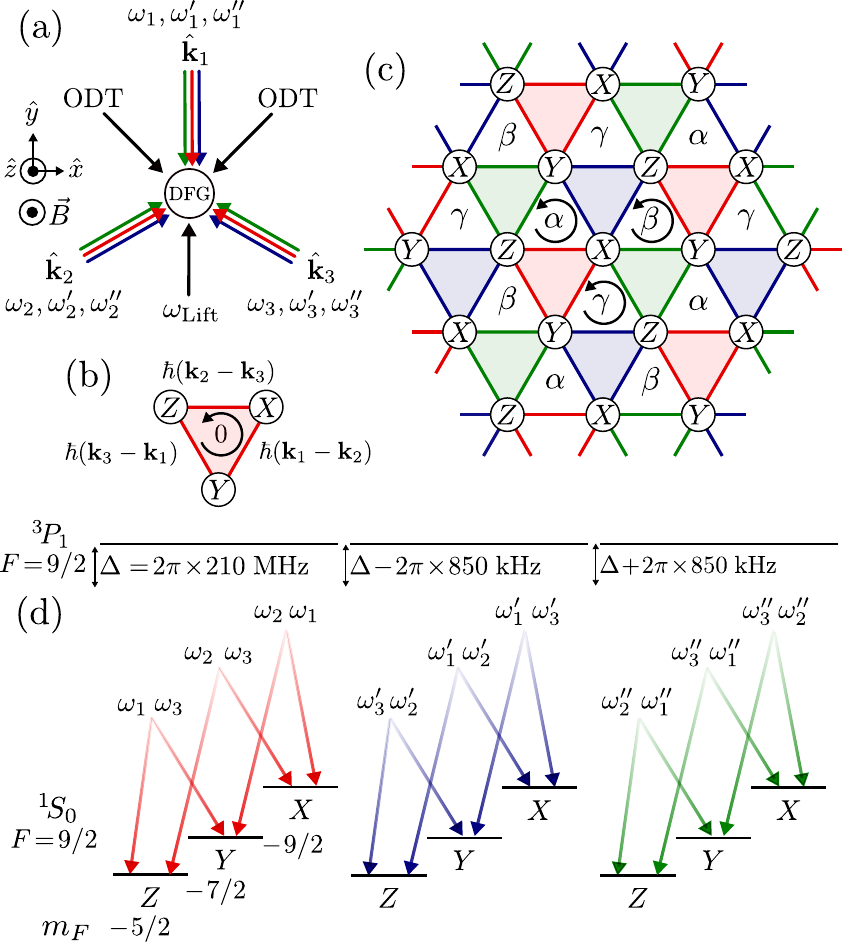}
\caption[Fig1]{ Experiment schematic and coupling details. (a) A  degenerate Fermi gas (DFG) in a crossed optical dipole trap (ODT) is exposed to cyclic Raman couplings between three internal spin states; see text for details. A beam $\omega_{\rm{Lift}}$ provides a nonlinear energy splitting between the states. 
(b) States $X,Y,Z$ are connected among themselves through spin-momentum exchange in units of $\hbar \mathbf{\delta k}_{ij} = \hbar ( \mathbf{k}_{i}-\mathbf{k}_{j})$. With a single set of beams, no net phase pickup is possible, denoted by the 0. (c) The couplings form a lattice in momentum space. Atoms encircling plaquettes labeled by $\alpha,\beta,\gamma$ can pick up a net phase. The link color indicates the frequency set in (d) to which the beams belong. 
(d) Details of the resonant couplings between the three internal states labeled $X,Y,$ and $Z$, which represent the nuclear angular momentum projections $m_F=-9/2,-7/2,-5/2$, respectively, in the ${}^1S_0(F=9/2)$ ground state. We circularly exchange the roles of the frequencies colored red ($\omega_i$) in the blue ($\omega_i'$) and green ($\omega_i''$) coupling sets. }

\label{SetupFigure}
\end{figure}

The setup implementing the optical couplings in Eqn. (1) is shown in Fig. 1(a). Up to three running-wave triplets of beams are incident on a degenerate Fermi gas (DFG) spin-polarized mostly into state $|X\rangle$ with $T/T_F= 0.36(5)$, where $T_F$ is the Fermi temperature. Each beam $\hat{\mathbf{k}}_i$ contains up to three frequencies $\omega_i,\omega_i',\omega_i''$, such that the energy difference between any two frequencies $\omega_i('(''))-\omega_j('(''))$ matches an energy difference in the $X,Y,Z$ manifold. These beams provide a Raman coupling between states, as in Fig 1(b). The quantization axis is defined by a $\hat{z}$-oriented magnetic field $B\approx 9.3~\rm{G}$, along which we align the linear polarization of a beam $\omega_{\rm{Lift}}$~\cite{Song2018,Han2019} providing a strong ac Stark shift that lifts the degeneracy of the states $X,Y,Z$. The coupling beam polarizations are linear and angled at $33(1)^\circ$ with respect to the $xy$-plane, projecting approximately equal intensity among the possible Raman transition types $\pi, \sigma^\pm$. When using all nine frequencies, the beams form an infinite lattice in spin-momentum space, as in Fig. 1(c). Nonzero gauge flux is possible only on upward-pointing triangles, corresponding to momentum transfers involving all three frequency sets. 

As shown in Fig. 1(d), the coupling beams utilize the dipole-forbidden transition $^1S_0(F=9/2) \rightarrow {}^3P_1(F=9/2)$, detuned below resonance by $\Delta/2\pi = 210~\rm{MHz}$. The transition's narrow linewidth $\Gamma/2\pi = 7.4~\mathrm{kHz}$ allows coherent manipulation with minimal spontaneous emission, and no significant destructive interference \cite{Naber2016,Zhang2019a} arises from the THz-separated fine structure states ${}^3P_0$ and ${}^3P_2$. In order to make each triplet unique, the upper-state detunings of the blue and green couplings are shifted by $\mp 37~E_R/\hbar$, larger than the $\approx 7.5~E_R/\hbar$ energy splittings. The role of the frequency $\omega_i$ in beam $\mathbf{k}_i$ is circularly rotated amongst the three triplets, such that all frequencies resonantly couple all spin states. 

In order to realize the lattice, careful attention must be paid to the energy levels of $X,Y,Z$, which are naturally degenerate. Since the ground states have $J\!=\!0$---rendering Zeeman shifts insignificant at our bias field---we use an  ac Stark shift approach. Further leveraging the narrow-line intercombination transition, the lift beam $\omega_{\mathrm{Lift}}$ is operated at $434.829943(5)~\rm{THz}$, midway between the hyperfine resonance lines $^1S_0(F=9/2) \rightarrow {}^3P_1(F=7/2)$ and $^1S_0(F=9/2) \rightarrow {}^3P_1(F=9/2)$. This light was designed to produce a strong tensor shift $\epsilon = 1.76 E_R/\hbar$ of the $X$ state, allowing each pair of Raman beams to uniquely couple two spin-momentum states. A necessary condition of the spin-momentum lattice model is that the coupling strengths $\Omega \ll \epsilon$. The coupling strengths here are $\Omega \approx 0.5 E_R/\hbar$ \cite{Supplemental}. 

\emph{Experimental sequence.} 
In our newly built apparatus, we source ${}^{87}\rm{Sr}$ from a commercial atomic oven from AOSense, which includes an integrated Zeeman slower and 2D MOT optics. After two MOT loading and cooling stages lasting $7 \rm{s}$~\cite{Stellmer2014,Snigirev2019}, the atoms are loaded into a crossed 1064~nm optical dipole trap (ODT) with an initial temperature of $\approx 2~\mu\mathrm{K}$. The vertical (horizontal) trapping frequency is ramped up to $2.160(5)~\rm{kHz}\; (313(1),397(2)~\rm{Hz}$), at which point we spin-polarize the sample with a series of pulses resonant with the different $m_F$ states via the $^1S_0(F=9/2) \rightarrow {}^3P_1(F=9/2)$ transition~\cite{Scazza2015, Whalen2019}. The ODT frequencies are then lowered back to $\approx\! 1$ kHz vertically, and forced evaporation proceeds over the next 10 s, finally reaching a quantum-degenerate sample. Without spin-polarization, we routinely achieve $T/T_F = 0.20$, where $T_F$ is the Fermi temperature, rising to $T/T_F = 0.36(5)$ when spin-polarized. Evaporation ends at mean geometric trap frequency $\overline{\omega} = (\omega_x \omega_y \omega_z)^{\frac{1}{3}} = 71.4(1)$~Hz, yielding a $50~\rm{nK}$ Fermi gas. Immediately following evaporation, the sample is spin-polarized in the state $X (80 \pm 7\%)$, and the ac Stark shifting beam $\omega_{\rm{Lift}}$ is ramped on in $0.5~\mathrm{ms}$. Via optical Stern-Gerlach imaging~\cite{Stellmer2011}, we verify that this timescale does not alter the spin polarization. We then introduce the coupling beams with a turn-on time of $<\! 1\mu\mathrm{s}$.

\begin{figure}[!t]
\includegraphics[width=\columnwidth]{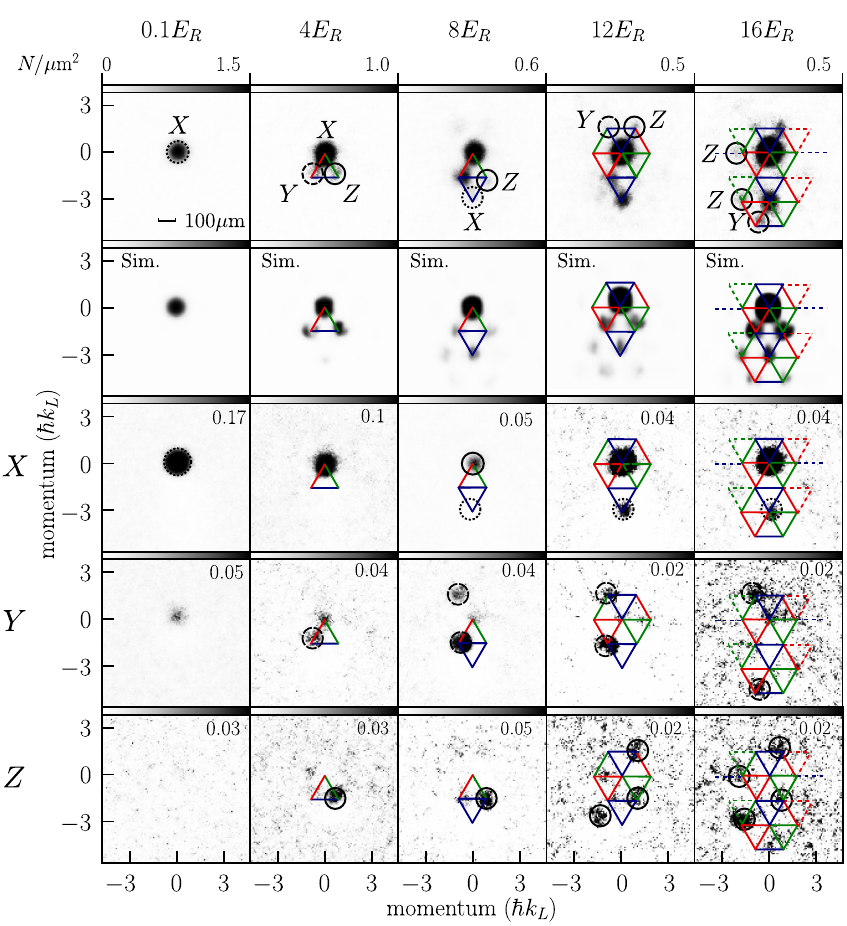} 
\caption{Demonstration of the spin-momentum lattice. A DFG is exposed to the nine-beam coupling scheme, with the $\mathbf{k}_1$ triplet frequency-swept in order to populate more lattice sites; see text for details. The sweep progress is indicated in units of recoil energy $E_R$. The top row shows spin-unresolved momentum-space images, in which the lattice is most clearly evident. The second row shows the predicted dynamics from simulations (Sim.). Subsequent rows show spin-resolved images. In each row, we draw a link connecting each lattice site with a color corresponding to the involved beams, as detailed in Fig. \ref{SetupFigure}(d). In the top row, we encircle and label each new lattice site as it becomes apparent during the sweep, and in the spin-resolved rows the expected populations are also encircled. As a guide to the eye, the links are also drawn across all rows. Each image is an average of $\approx 100$ experimental runs taken at 12~ms time of flight. The peak atomic density is indicated in the top right of each image, in units of $N$ atoms per $\mu\mathrm{m}^2$.}
\label{LatticeFigure}
\end{figure}

We demonstrate the spin-momentum lattice in Fig. \ref{LatticeFigure}. Since the average starting atomic momentum $\langle p\rangle \ll 4 \hbar k_R$, the fermions initially occupy only a small spread of states in  $|X,\mathbf{q}\approx 0 \rangle$, where $\mathbf{q}$ is the quasimomentum. In order to fill more sites, we emulate motion along a single dimension by subjecting the atoms to an inertial force~\cite{Chalopin2020} along $\hat{\mathbf{k}}_1$, ramping all three of that beam's frequencies at a rate $\hbar \partial_t (\omega_1,\omega_1',\omega_1'') = 16.607~E_R/\rm{ms}$\footnote{We do not expect this sweep rate to be adiabatic with respect to the current Rabi coupling strengths.}. Hopping to neighboring sites is made favorable when the frequency difference between two coupling beams matches the energy and recoil shifts between states, providing enhanced state transfer between initial state $|X,\mathbf{q}\rangle$ and $|X,\mathbf{q} - \mathbf{K} \rangle,|Y,\mathbf{q}- \mathbf{K}\rangle,|Z,\mathbf{q}- \mathbf{K}\rangle$ for some reciprocal lattice vectors $\mathbf{K} = n_1 \mathbf{q}_1+n_2 \mathbf{q}_2$ with integers $n_1,n_2$. After a varying sweep time, all optical fields are quenched off, releasing the atoms from the harmonic trap. Atoms that have tunneled to different lattice sites acquire a concomitant increase in momentum, in discrete units of the two-photon Raman momentum  $\hbar k_R = \sqrt{3}/2 \hbar k_L$. The lattice sites become spatially resolved after 12~ms time of flight, since the starting momentum distribution's full-width half-max width is $1.05(1) \hbar k_L$ and external heating by spontaneous emission from ${}^3P_1$ is minimal~\cite{Supplemental}. The atoms are then absorption-imaged in the $xy$ plane using the ${}^1S_0\rightarrow{}^1P_1$ transition at 461~nm ($\Gamma_{461}/2\pi = 30 \mathrm{MHz}$), which images all spins with approximately equal efficiency~\cite{Barker2016}. The bias magnetic field is kept on at all times, in order to maintain the spin quantization axis.

The individual columns of Fig. \ref{LatticeFigure} demonstrate spin- and momentum-resolved imaging at various quench times. Sweep time is indicated by the final frequency deviation of the swept beam, in units of $E_R$. Intuitively, one would not expect stationary atoms ($\langle p \rangle \approx 0$) to tunnel before at least overcoming the recoil shift $4E_R$, and we observe this in the experiment. Denoting transferred momentum by $\hbar \delta k_{ij} = \hbar(|\mathbf{k}_i-\mathbf{k}_j|)$, and referring to beam triplets by their colors in Fig. 1(d), when the sweep reaches $4E_R$ we see beams $\omega_1,\omega_2$ from the red triplet driving the corresponding Raman transition $|X,0\rangle \rightarrow |Y,\hbar \delta k_{12} \rangle$; similarly, the green beams $\omega_1',\omega_3'$ allow $|X,0\rangle \rightarrow |Z,\hbar \delta k_{13} \rangle$. By $8E_R$, atoms have firmly populated sites $|Y,\hbar  \delta k_{12} \rangle ,|Z,\hbar \delta k_{13} \rangle$, with initially-faster-moving atoms beginning to populate the site $|X,p=-3 \hbar k_L\rangle$, completing a traversal of the first Brillouin zone. By $12E_R$, more atoms have tunneled through the Brillouin zone, and the momentum center-of-mass proceeds downward at $16 E_R$; imaging becomes increasingly difficult due to the lower atom density, so we terminate here. As a consistency check, we also demonstrate spin-resolved imaging using spin blasts \cite{Song2016,Supplemental} in order to verify that sites on the SM lattice are of the expected spin projection, $m_F$. We observe good consistency with the SM lattice model as drawn in Fig. 1(d), although mechanical effects of the spin-blasts can mask some lattice sites; notably, the $X$-site at $\mathbf{p}=-3\hbar k_L \hat{y}$ (see \cite{Supplemental}). Our interaction model shows qualitative agreement with the data for a scaled value of the measured Rabi coupling strengths. Some disagreement is evident, especially at the lattice sites with momenta $\mathbf{p}=\sqrt{3}/2 \hbar k_L$, which are predicted to have a stronger amplitude than is observed. We attribute these mismatches to off-resonant effects not included in our effective Hamiltonian \cite{Supplemental}. 

\begin{figure}[t]
\includegraphics[width=\columnwidth]{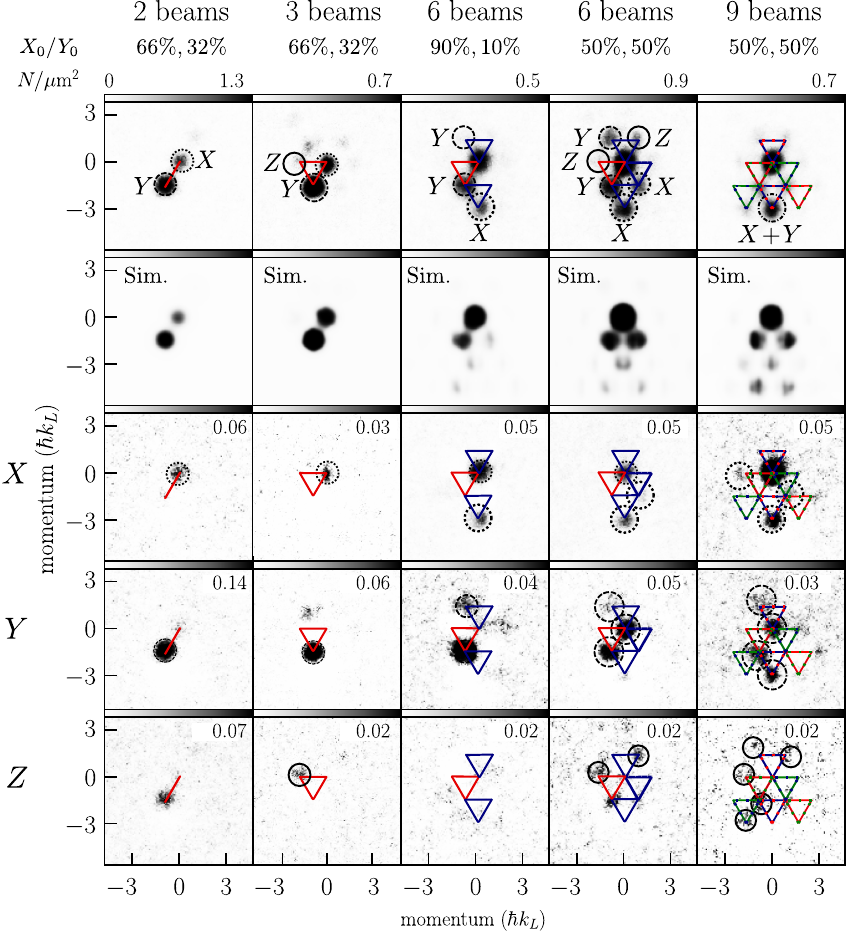}
\caption{Building the spin-momentum lattice; all images taken at a common sweep time of $12 E_R$. The drawn links and circles are as described in Fig.~2, with the second row showing the predicted dynamics (Sim.). Each removed frequency corresponds to a missing link in the full lattice setup, allowing the exploration of quasi-1D SOC in the first column, to 2D SOC in the second column, culminating in the full lattice shown in the last column. The multiply-colored links in the last two columns indicate two SM lattices simultaneously overlaid: a two-state configuration, in which the experiment began with equal populations of $X,Y$. }
\label{BrokenLinksFigure}
\end{figure}

The lattice scheme presented here is readily tunable. Although the full spin-momentum lattice is composed of nine frequencies, we can remove links between lattice sites at will. We explore this flexibility in Fig. \ref{BrokenLinksFigure}, where we show the driven dynamics experiment of Fig. \ref{LatticeFigure}, but now with all images taken at a common sweep time $12E_R$. In the two-beam scheme, composed of a single frequency in each of two beams $\hat{k}_1,\hat{k}_2$, we have reduced the system to a 1D SOC model between an effective spin up ${|\!\uparrow\rangle} = |X,\mathbf{q}\rangle$ and spin down ${|\!\downarrow \rangle} = |Y, \mathbf{q} \rangle$~\cite{Wang2012,Song2016}. In the 3-beam case, with a single frequency in each of the $\hat{k}_1,\hat{k}_2, \hat{k}_3$ beams, we have a 2D spin-orbit coupling~\cite{Campbell2011,Campbell2016,ValdesCuriel2021} cyclically linking the three states $X,Y,Z$. The 6-beam case consists of beams $\hat{k}_1,\hat{k}_2, \hat{k}_3$ each possessing two frequencies, labeled by their colors red and blue as labeled in Fig. \ref{SetupFigure}. In the last two columns, we explore the dynamics starting from an even spin mix of states $X$ and $Y$, which, in the 9-beam experiment, can be visualized as two SM lattices overlapped on $\mathbf{p}=0$. 

\emph{Conclusion and outlook.} We have demonstrated a two-dimensional fermionic spin-momentum lattice without the use of standing waves. This adds to the wealth of cold atom synthetic dimension platforms available to study topological materials. The system's 15~ms lifetime exceeds our current experimental duration by a factor of 10, and could be further improved with increased Raman detuning $\Delta$. The current lift beam strength imposes a 30~ms limit, which can be relaxed under appropriate conditions~\cite{Supplemental}. The number of visible lattice sites can be increased with larger Rabi coupling strengths, or by slowing the sweep rate, which would couple more atoms out of the $p\approx0$ momentum class. The spin-resolved imaging presented here could be improved by using stronger blast pulses to overcome the Doppler shifts among the lattice's numerous momentum states.

This work launches a novel platform for exploring topological physics with optical flux lattices. The natural extension of this work would be to load the atoms adiabatically into the lowest band and set nonzero coupling phases such that a gauge flux appears on the plaquettes labeled $\alpha,\beta,\gamma$ in Fig. \ref{SetupFigure}. The topology of the band structure could then be probed using established anomalous velocity techniques~\cite{Price2012,Aidelsburger2014,Chalopin2020}, which involve accelerating the dressed atoms in the same manner as done here. Demonstrating this topology would enable the exploration of many-body fractional Hall states~\cite{Cooper2013}.\\ \\

\begin{acknowledgments}
We would like to thank C. Yu for assistance in constructing an early version of the experiment. We thank S. Mossman for helpful discussions of the experimental design, and are grateful for helpful comments from I. Spielman. 
\end{acknowledgments}
\bibliography{references}

\clearpage

\section*{Supplementary material}
\subsection{Model.}

	\begin{figure}[htbp]
		
		\includegraphics[width=0.9\columnwidth]{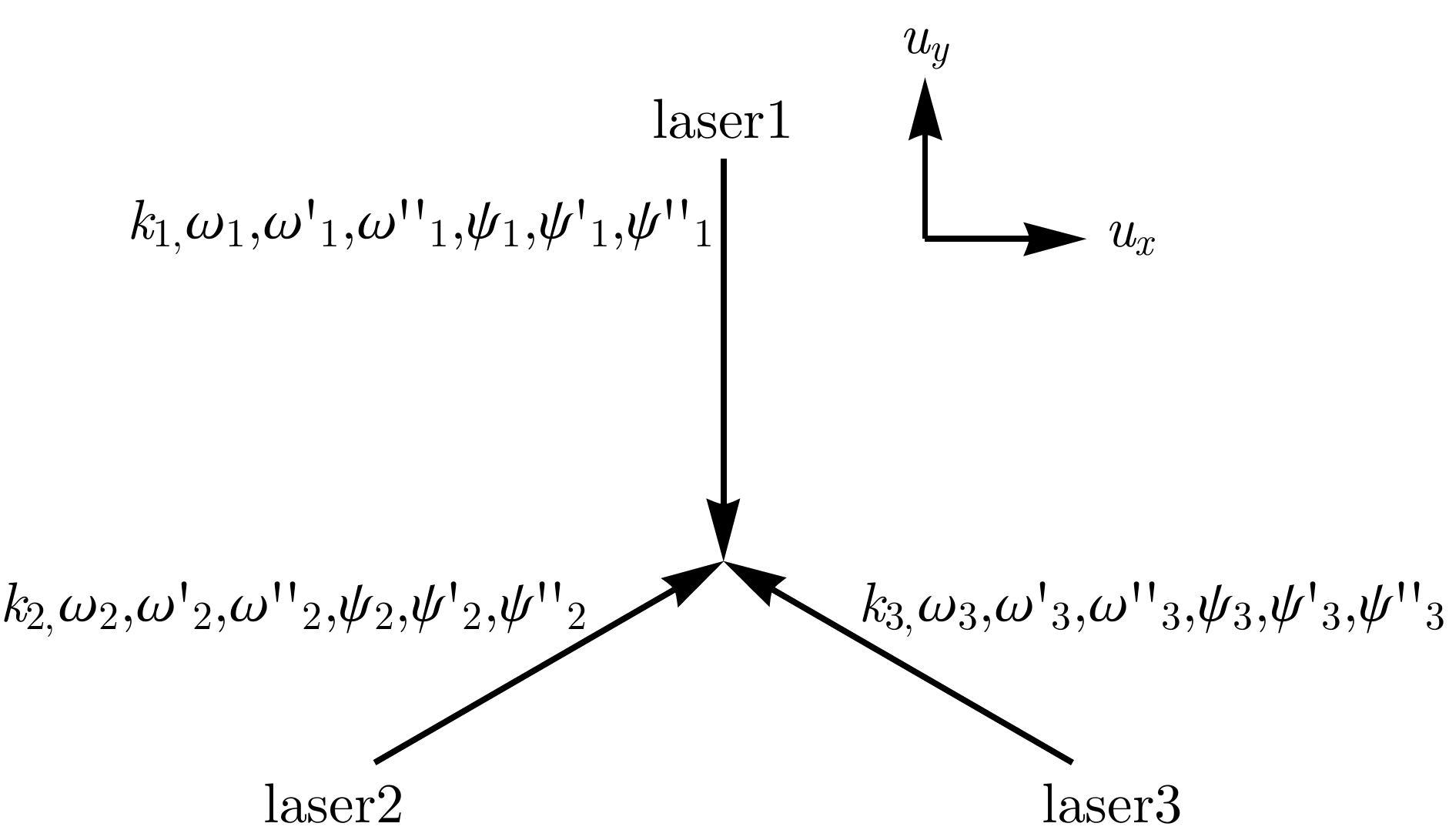}
		\caption{}
		\label{fig_Exp}
		
	\end{figure}
	
The spin-momentum (SM) lattice setup is shown in \figref{fig_Exp}. We define the wave vectors as
	$$\mathbf{k}_{1} = -k_L\hat{u}_{y},\hspace{0.05in} \mathbf{k}_{2} = \frac{k_L}{2}(\sqrt{3}\hat{u}_{x}+\hat{u}_{y}), \hspace{0.05in} 
		\mathbf{k}_{3} = \frac{k_L}{2}(-\sqrt{3}\hat{u}_{x}+\hat{u}_{y})$$
where $k_L=2\pi/\lambda.$ Note that although these beams do not have the same frequency, the difference between wavenumbers is negligible. Since the transitions induced by these lasers involve two-photon processes, the corresponding momentum transfer is related to differences between $\mathbf{k}_1,\mathbf{k}_2,\mathbf{k}_3$. We define the single-photon recoil vector $k_{R}=2\pi/\lambda \sin(\theta/2)$ where $\theta=120^\circ$ is the angle between any two beams. The recoil vector becomes $k_{R} = \sqrt{3}\pi/\lambda = \sqrt{3}k_L/2$, so the wavevectors become
	$\mathbf{k}_{1} = -\frac{2}{\sqrt{3}}k_{R}\hat{u}_{y},\hspace{0.1in} \mathbf{k}_{2} = \frac{k_{R}}{\sqrt{3}}(\sqrt{3}\hat{u}_{x}+\hat{u}_{y}), \hspace{0.1in} 
		\mathbf{k}_{3} = \frac{k_{R}}{\sqrt{3}}(-\sqrt{3}\hat{u}_{x}+\hat{u}_{y}).$
The relative momentum $\mathbf{q}_{i}$ is defined as
	$\mathbf{q}_{1} = \mathbf{k}_{1}-\mathbf{k}_{2} = k_{R}(-\hat{u}_{x}-\sqrt{3}\hat{u}_{y}),\hspace{0.1in} \mathbf{q}_{2}=\mathbf{k}_{2}-\mathbf{k}_{3} = 2k_{R}\hat{u}_{x},\hspace{0.1in}\mathbf{q}_{3} = \mathbf{k}_{3}-\mathbf{k}_{1} = k_{R}(-\hat{u}_{x}+\sqrt{3}\hat{u}_{y}).$
All magnitudes are $2k_{R}$, consistent with the motivation of defining the single-photon recoil vector. For convenience, we also define $\omega_{Y\rightarrow X} = \omega_{1}-\omega_{2} = \omega''_{2}-\omega''_{3}=
	\omega'_{3}-\omega'_{1},
	\omega_{X\rightarrow Z} = \omega''_{1}(t)-\omega''_{2} = \omega'_{2}-\omega'_{3}=
	\omega_{3}-\omega_{1},
		\omega_{Z\rightarrow Y} = \omega''_{1}-\omega''_{2} = \omega'_{2}-\omega'_{3}=
	\omega_{3}-\omega_{1}(t)$.  The time dependence only shows up in $\omega_{1},\omega'_{1},\omega''_{1}$ since only the frequencies in beam $\mathbf{k}_1$ are swept during the experiment.

To simulate the quantum dynamics in the spin-momentum (SM) lattice, we first set up the Hamiltonian in the Bloch basis, $\psi_{\mathbf{q},\alpha}(\mathbf{r})=e^{i\mathbf{q}\cdot \mathbf{r}}u_{\mathbf{q},\alpha}(\mathbf{r})$ where $\mathbf{q}$ is the crystal momentum, $\alpha$ is the spin species and $u_{\mathbf{q}}$ is a periodic function. The corresponding Schr\"{o}dinger equation can be expressed as
$$\bigg(\frac{(\hat{\mathbf{p}}+\hbar \mathbf{q})^{2}}{2m}+\epsilon_{\alpha}+\Omega(\hat{\mathbf{r}},t)\bigg)u_{\mathbf{q},\alpha}(\mathbf{r},t)=i\hbar \frac{\partial}{\partial t}e^{i\mathbf{q}\cdot \mathbf{r}}u_{\mathbf{q},\alpha}(\mathbf{r},t)$$
Note that $\epsilon_{\alpha}$ labels the energy of the internal state. Due to the periodicity of the $u_{\mathbf{q,\alpha}(\mathbf{r},t)}$, we can  decompose $u_{\mathbf{q,\alpha}(\mathbf{r},t)}$ into its Fourier components,
	$$u_{\mathbf{q},\alpha}(\mathbf{r},t) = \sum_{\mathbf{K}}c_{\mathbf{K},\alpha}(\mathbf{q},t)e^{i\mathbf{K}\cdot \mathbf{r}}$$
where $\mathbf{K}= n_{1}\mathbf{q}_{1}+n_{2}\mathbf{q}_{2}$ is the reciprocal lattice vector, described by integers $n_1,n_2$. The Schr\"{o}dinger equation becomes
	\begin{equation*}
	\begin{split}
		\label{eq:TDSE}
		&\sum_{\mathbf{K}}e^{i\mathbf{K}\cdot\mathbf{r}}\bigg(\frac{\hbar^{2}(\mathbf{q}+\mathbf{K})^{2}}{2m}+\epsilon_{\alpha}+\Omega(\hat{\mathbf{r}},t)\bigg)c_{\mathbf{K},\alpha}(\mathbf{q},t) \\
		&= 
		\sum_{\mathbf{K}}e^{i\mathbf{K}}
		i\hbar \frac{\partial}{\partial t}c_{\mathbf{K},\alpha}(\mathbf{q},t) 
		\end{split}
		\end{equation*}
	Next, we formulate the interaction term with frequency sweep velocity $v$ as
	\begin{equation*}
	\begin{split}
		\Omega(\hat{\mathbf{r}},t) &= \sum_{n}\sum_{\alpha\neq\beta} \Omega_{\alpha\beta}e^{i \mathbf{q}_{n}\cdot \mathbf{r}-i(\delta_{n,1}-\delta_{n,3})vt^{2}}e^{-i \omega_{\beta\rightarrow\alpha} t}\ket{\alpha}\bra{\beta}\\
		&= \sum_{n}\sum_{\alpha\neq\beta} \Omega_{\alpha'\beta'}e^{i \mathbf{q}_{n}\cdot \mathbf{r}-i(\delta_{n,1}-\delta_{n,3})vt^{2}}\ket{\alpha'}\bra{\beta'} 
	\end{split}
	\end{equation*}
where $\delta_{m,n}$ is the Kronecker delta, and the last equation comes from using the rotating frame to absorb the usual oscillation term. 

In our simulation, we use a radial grid of 832 points in momentum space and use the LSODA differential equation solver \cite{Petzold1983} to predict the dynamics at each point, time-evolving a Gaussian distribution which closely matches the initial momentum spread. We fix the 9 interaction strengths according to Table \ref{RabiTable}, the values there obtained as described in Section \ref{CouplingStrengths}. We have empirically found that these interaction strengths are approximately 20\% too large to account for the observed dynamics, and thus uniformly scale all Rabi strengths accordingly. 

\begin{figure}[ht]
\includegraphics[width=\columnwidth]{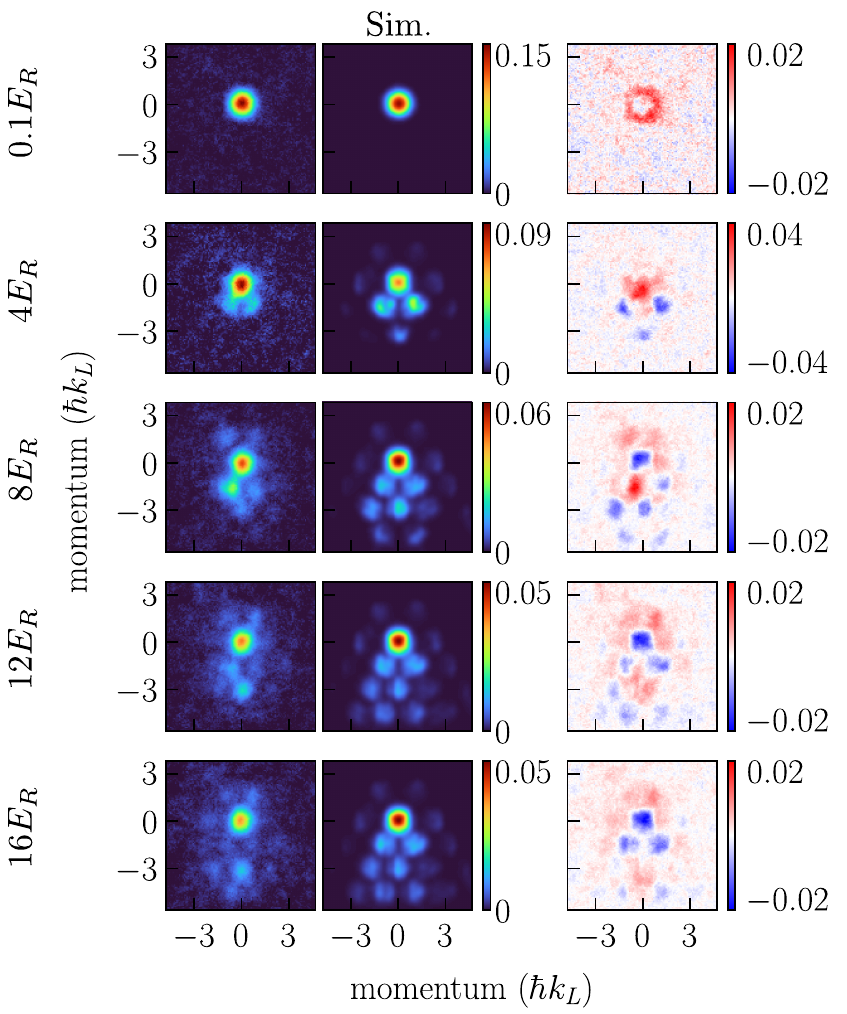}
\caption{Comparison of the 9-beam SM lattice data from main text Fig. 1 (left column) with the model (middle column), showing the difference image (right column). The model time-evolves a Gaussian momentum distribution with starting width $1.05 \hbar k_L$. The coupling strengths used are given in Table \ref{RabiTable}, all of which are uniformly scaled down by $20\%$. We apply to the model images a small Gaussian blur (std. dev. = 2 pixels) commensurate with the imaging resolution of $\approx 1.5 $ pixels = $0.1 \hbar k_L$. The units are optical density.}
\label{ModelFigure}
\end{figure}

\begin{figure}[t]
\includegraphics[width=\columnwidth]{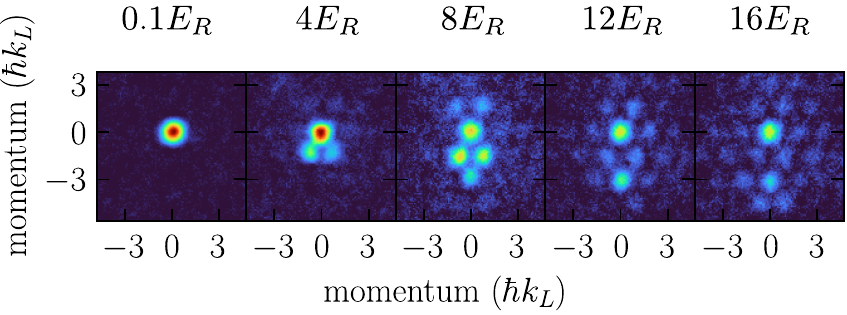}
\caption{Impact of too-small triplet frequency separation. An earlier version of this experiment separated the triplets by only $\pm 525$ kHz (now $\pm 850$ kHz), resulting in the strong population of positive-momentum lattice sites, a result not expected from the direction of the frequency sweep. }
\label{XCoupling}
\end{figure}

Still, some disagreement with the model persist, as can be seen in Fig. \ref{ModelFigure}. The model under-predicts the upward-going atomic motion, an effect responsible for the over-predicted population with $\mathbf{p}\approx 0$ at later  evolution times. These effects were especially noticeable in an earlier version of the experiment, in which we had a smaller triplet separation; compare with Fig. \ref{XCoupling}, which shows significant population in modes with $\mathbf{p}>0$ due to cross-coupling between triplet beams.

We attribute the discrepancy to both the lingering effects of inter-triplet interference, and to off-resonant Raman couplings in the SM lattice---i.e., couplings which reverse the notion of which beam is considered `pump' and which is considered `Stokes.' By increasing the tensor shift and separation between frequency triplets (currently $\pm 850$ kHz), such effects can be further suppressed. 

\subsection{Light sources.} 

The ground state degeneracy is lifted by the tensor shift induced by $\omega_{\rm{Lift}}$. For this beam, we source $182(2)~\rm{mW}$ of light from a mode-locked commercial Ti:Sapph laser (M Squared SolsTis XS). The Raman coupling beams are sourced from a commercial amplified diode laser (Toptica Photonics AG)  locked to an external cavity (Stable Laser Systems) with linewidth $<\!\! 1000$~Hz. At the experiment, the Raman coupling beams are split from the ``stirring" beam~\cite{Mukaiyama2003,Snigirev2019} path, which is used during the MOT and spin polarization phases. They pass through a +290~MHz shifting acousto-optic modulator (AOM) about 1~m from the atoms, which spatially filters out any crystal-scattered 0th-order light that would cause heating. This light is then split to three independent AOMs, which shift by -80 MHz and imprint the Raman-resonant frequencies, yielding an overall detuning $\Delta/2\pi= 210$~MHz from ${}^3P_1(F=9/2)$. One beam ($\mathbf{k}_3$) is fiber-coupled through a polarization-maintaining fiber, and the other two ($\mathbf{k}_1,\mathbf{k}_2$) are free-space. By placing the free-space beam shaping optics after the AOMs and using relatively large beam waists of $\approx 200 \mu \mathrm{m}$, we constrain beam misalignment (from AOM deflection angle spread amongst the 3 frequencies) to effect a $<5$\% change on the associated Rabi frequencies. The same arbitrary waveform generator (Spectrum Instrumentation M4i662-x8) drives all three Raman AOMs and is phase-locked to a 10 MHz Rubidium reference standard (Stanford Research Systems FS725). We do not phase lock any lasers.

\subsection{Coupling strengths.}\label{CouplingStrengths}

\emph{Rabi oscillations and lattice closure.} By observing Rabi oscillations between pairs of coupling beams, we calibrate the coupling strengths $\Omega_i$ and quantify the degree to which the SM lattice is limited to the spin projections $m_F=-5/2,-7/2,-9/2$. The Rabi measurement protocol differs from the main experiment, due to our dual use of the Ti:Sapph laser as both lift beam and as a source of optical Stern-Gerlach (OSG) pulses, which are separated by several GHz. While we could bridge the difference with a series of AOMs, we instead apply an open-loop drive voltage to the laser's cavity lock in order to change the laser's function from lift to OSG, as follows. 

\begin{figure}[h]
\includegraphics[width=\columnwidth]{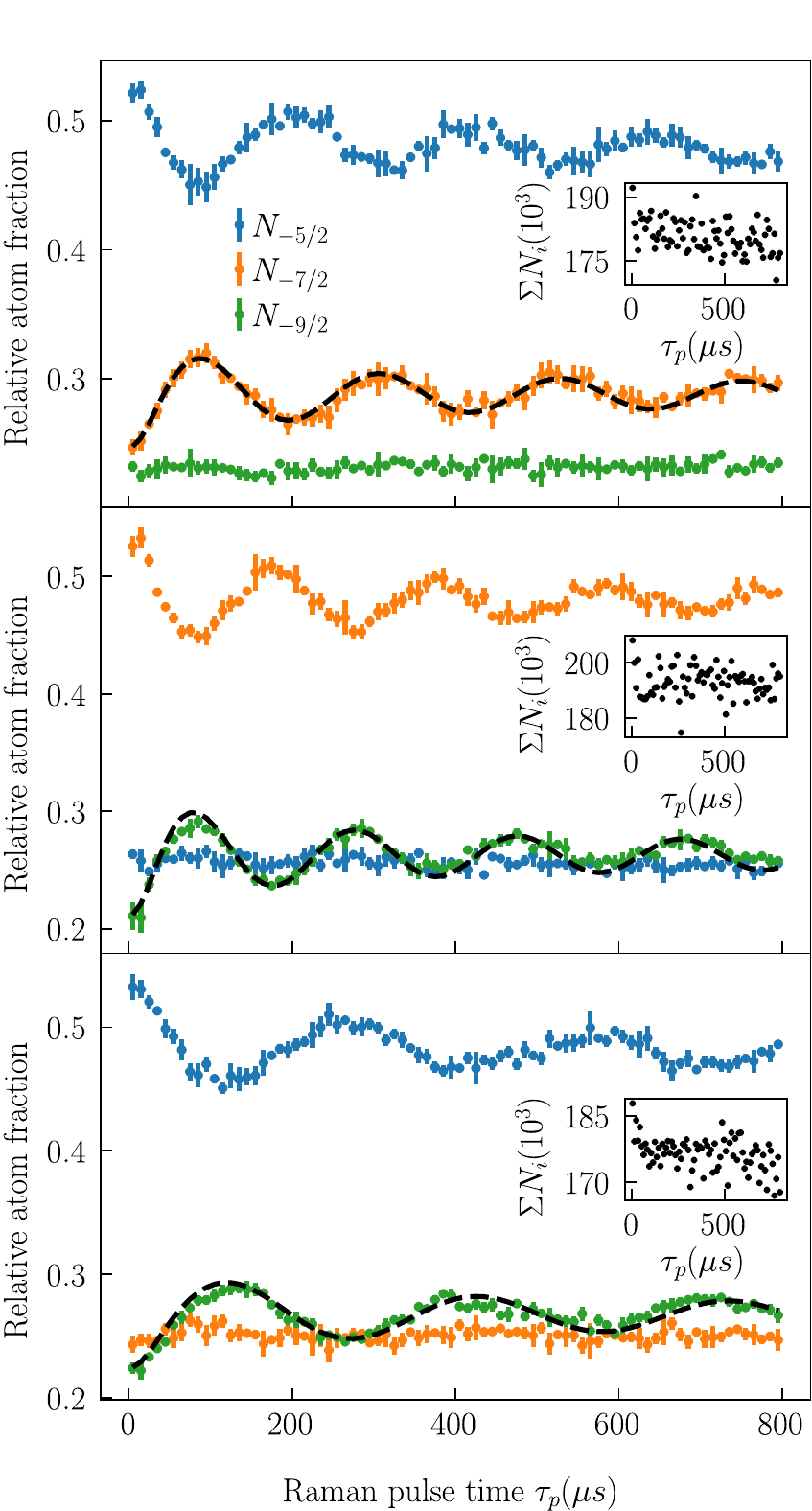}
\caption{ Coupling strength calibration and lattice closure measurement from OSG data. Rabi oscillations due to a pulse resonant with (top) $-5/2\rightarrow -7/2$ (middle)$-7/2\rightarrow -9/2$ (bottom) $-5/2\rightarrow -9/2$. Insets show total atom count in the $XYZ$ triplet, $\Sigma_{i}^{-5/2,-7/2,-9/2}N_i$, at each pulse length $\tau_p$. Triplet losses due to coupling to adjacent $m_F$ states such as $-1/2, -3/2$ would present as a decrease in overall atom count, bounded here to $<10$\%. }
\label{RabiFigure}
\end{figure}

After applying a Raman pulse of varying duration, both Raman and lift beams are snapped off while the ODT is kept on. In the next 50 ms, we ramp the laser frequency from the lift beam frequency $434.829943(5)~\rm{THz}$ to the OSG frequency $434.828370(30)$ THz. The larger frequency uncertainty induced by the open-loop control translates to small positional variations of the atomic spin populations from shot-to-shot. We account for this in the data analysis by binning the spin locations relative to the location of the locally-maximum atom cloud, which is unambiguously the ``spin-up" population since the transfer efficiency is never larger than 0.5 (see Fig. \ref{RabiFigure}). Empirically, the 50 ms ramp time dampens the Raman-kicked atomic motion sufficiently that the OSG pulse separates the $m_F$ states along the $\hat{y}$ direction only.

We summarize these measurements in Fig. \ref{RabiFigure}. Atoms with quasimomentum $\mathbf{q}$ predominantly populate  the ``spin up" state $|\!\uparrow  = -5/2,\mathbf{q}\rangle$ in (a) and (c), or $|\!\uparrow = -7/2,\mathbf{q}\rangle $ in (b). A light pulse of varying length is shone onto the tensor-shifted atoms, the OSG beam is fired as described above, and then all coupling and trapping beams  are snapped off. This results in Rabi oscillations between bare states $|\!\uparrow, \mathbf{q}\rangle$ and $|\!\downarrow, \mathbf{q}+\sqrt{3} \hbar k_L\rangle$. We fit to the spin-down population with a model ~\cite{Burdick2016,Hu2019}

\begin{equation}
P_{\downarrow} = A\int_{-v_0}^{v_0} dv P_e(-\sqrt{3} k_L v,\Omega,t)G_v(\sigma) + N_{\downarrow,0}
\end{equation}
where $A$ is the amplitude, $t$ is the pulse length, $P_e(\delta,\Omega,t)=\Omega^2/(\Omega^2+\delta^2)\sin^2(t/2\sqrt{\Omega^2+\delta^2})$, $G_v(\sigma)$ is the Gaussian velocity distribution with standard deviation $\sigma = \sqrt{2 k_B T/m}$, and $N_{\downarrow,0}$ is the fixed amount of atoms populating $|\!\downarrow \rangle $ at $t=0$. With only two fit parameters, $\Omega$ and $A$, we find good agreement with the model, observing the decoherence characteristic of finite-temperature SO-coupled fermions~\cite{Wang2012,Song2016}. The Rabi coupling strengths between each link in the spin-momentum lattice are obtained by scaling the Rabi frequencies obtained in Fig. \ref{RabiFigure}. The scaling considers both the respective branching ratios of each transition, and also the small measured differences in power between each tone in the driving AOMs. The strengths are listed in Table \ref{RabiTable}, with a total uncertainty considering the involved beam waists (measured through Fig. \ref{RabiFigure}), polarization angle uncertainty of $1^\circ$, and the measured power uncertainty in each tone, which we take to be 5\%. 

The insets of Fig. \ref{RabiFigure} indicate the total atom number in all three spin states $X,Y,Z$. Since the timescale of spontaneous emission (see Section \ref{lifetime}) is small compared to the pulse durations, atom loss is due to atoms coupling to the adjacent dipole-allowed states $m_F=-1/2,-3/2$, which are not counted in the OSG images. The loss into these external states is expected to scale with the Rabi coupling strength and here is at most 10\%, showing that the spin-momentum lattice experiment is closed to the spin states $-9/2,-7/2,-5/2$. 

\begin{figure}[!h]
\includegraphics[width=0.65\columnwidth]{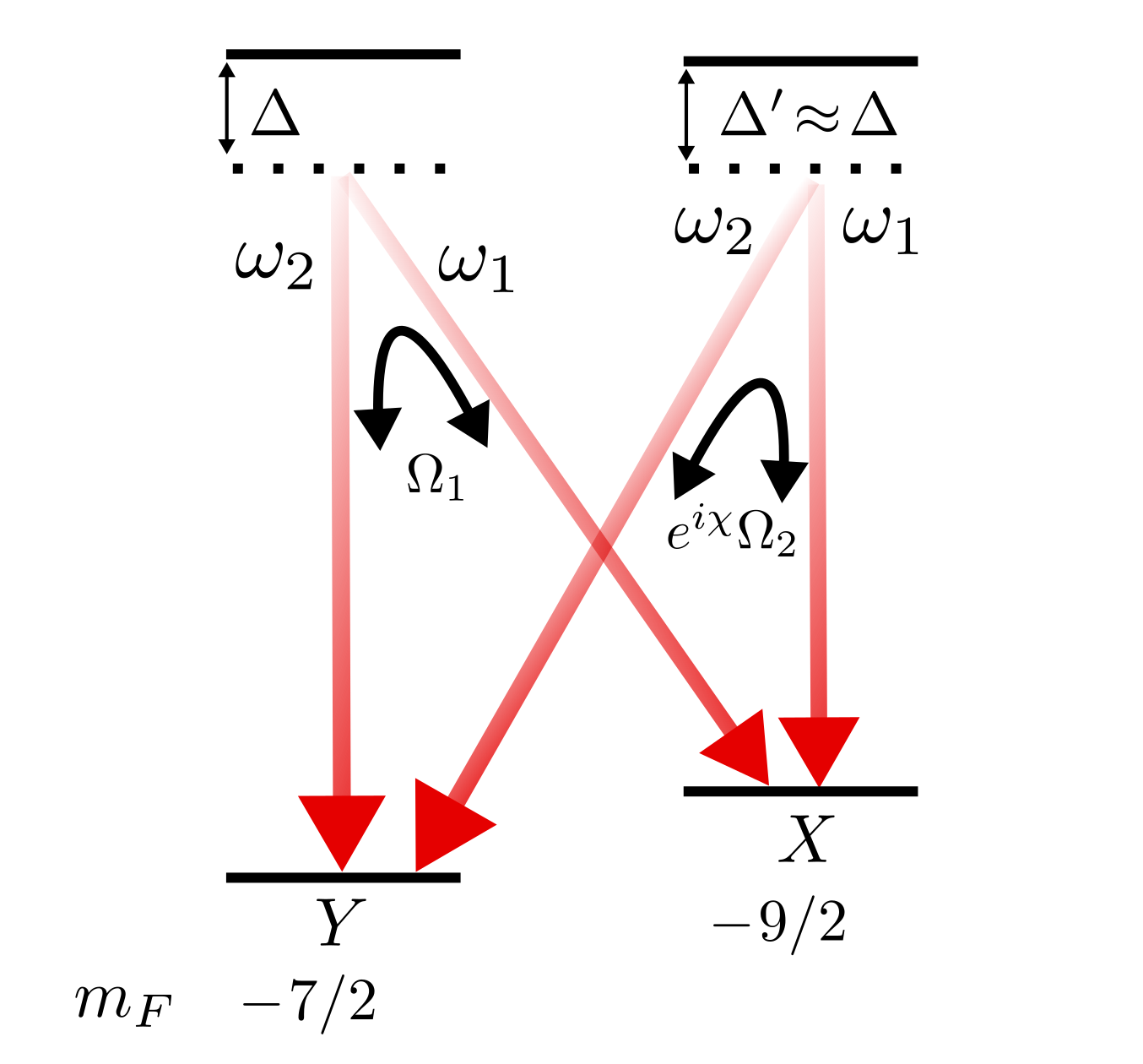}
\caption{Raman coupling in the double-$\Lambda$ configuration. States $X$ and $Y$ (angular momentum projections $m_F=-9/2$ and $m_F=-7/2$, respectively) are coupled by two distinct two-photon transitions driven by beams with the same frequency but differing polarization. The resulting effective coupling strength depends on the interference between $\Omega_1$ and $\Omega_2$. In the SM lattice, a similar double-$\Lambda$ configuration exists between states $Z (m_F=-5/2)$ and $Y (m_F=-7/2)$. }
\label{RabiPhaseFigure}
\end{figure}

\emph{Double-$\Lambda$ Raman interference.}\label{RamanInterference} In describing the net effective Rabi strengths $\Omega_{\mathrm{Eff}}$, which form each link in the lattice, we must carefully consider the polarizations. Because each beam $\mathbf{k}_i$ can drive $\sigma^+,\sigma^-$, and $\pi$ transitions, ``double-$\Lambda$" type couplings \cite{Wang2018} are realized between the links $X \leftrightarrow  Y$ and $Y \leftrightarrow  Z$; see Fig. \ref{RabiPhaseFigure} for the specific coupling between $X$ and $Y$. The coupling frequencies may cause a $\pi \sigma^-$ or $\sigma^+ \pi$ transition, since the upper-state detunings in each case satisfy $\Delta/\Delta' \approx 1$ and thus have similar coupling strengths (in contrast, the couplings $X \leftrightarrow Z$ may proceed only via  $\sigma\sigma$ due to dipole selection rules). For such links, we define a net effective coupling strength $\Omega_{\mathrm{Eff}} = \Omega_1+e^{i\chi} \Omega_2$, where $\chi$ is the relative phase and $\Omega_i$ is the two-photon Rabi frequency associated with a specific polarization scheme, as in \cite{Wu2016,Sun2018}. In our experiment, the phase between $\pi$- and $\sigma$-components is fixed, unlike experiments in Refs. \cite{Wu2016} and \cite{Sun2018}, which modify this phase using a tunable path length difference or electro-optic modulator, respectively. As we show, our phases ${\chi}$ are fixed by the beam intersection angle. 

Consider $\omega_1$ inside beam $\mathbf{k}_1$, propagating along the $-\hat{y}$ direction as in the main text Fig. \ref{fig_Exp}, with polarization angle $\theta=33^\circ$ with respect to the $xy$-plane. The electric field, with real amplitude $E_{10}$, including a possible phase $\psi$, which can vary from shot to shot, is  

\begin{equation}\label{E1Field}
\mathbf{E}_1 = E_{10}  (\cos(\theta) \hat{x} + \sin(\theta) \hat{z}) e^{ i \mathbf{k}_1 \cdot \mathbf{r}} e^{i\psi}
\end{equation}

\noindent Similarly, the electric field describing $\omega_2$ in $\mathbf{k}_2$ is 

\begin{equation}\label{E2Field}
\begin{split}
\mathbf{E}_2  =  E_{20}  (& -\sin(\phi) \cos(\theta) \hat{x} + \cos(\phi)\cos(\theta) \hat{y} + \\ 
& \sin(\theta) \hat{z}) e^{ i\mathbf{k}_2\cdot \mathbf{r}}
\end{split}
\end{equation}
 
\noindent with $\phi = 30^\circ$ being the acute angle between $\hat{k}_2$ and the $x$-axis. These two fields drive a Raman transition $X\leftrightarrow Y$ when their frequency differences are close to the energy splittings. As shown in Fig. \ref{RabiPhaseFigure}, there are two ways an atom can make the jump, with respective two-photon Rabi frequencies $\Omega_1,\Omega_2$ proportional to $E_1^{\pi*} E_2^{\sigma-}$,$E_1^{\sigma+*} E_2^{\pi}$. The $\pi$-components of the fields are those along the $\hat{e}_z$ direction, and the $\sigma^{\pm}$ components are those along the directions $\hat{e}_\pm =\mp 1/\sqrt{2}(\hat{x}\pm i\hat{y})$. We also note that the geometric scaling factor resulting from the branching ratio associated with $\Omega_1$ is $-7\sqrt{2}/99 = \exp(i\pi) 7\sqrt{2}/99$ (associated with $\Omega_2$ is $\sqrt{2}/11$). We then find 

\begin{equation}
\begin{split}
\Omega_1 \propto E_1^{\pi*} E_2^{\sigma-} =  & E_{10}  E_{20}/\sqrt{2} \sin(\theta) \cos(\theta) e^{-i\psi} \\
& \times e^{i(\mathbf{k}_2 - \mathbf{k}_1) \cdot \mathbf{r} } e^{i (-\phi-\pi/2)}  
\end{split}
\end{equation}
\begin{equation}
\begin{split}
\Omega_2 \propto E_1^{\sigma+*} E_2^{\pi}= & E_{10}  E_{20}/\sqrt{2} \sin(\theta) \cos(\theta) e^{-i\psi} \\
&\times e^{ i(\mathbf{k}_2 - \mathbf{k}_1)\cdot \mathbf{r}}
\end{split}
\end{equation}

\noindent  We find the total phase difference $\chi = -\phi-\pi/2 = -2\pi/3$ between these two paths; we furthermore see that any phase fluctuation in $\psi$ is common-mode and so does not affect $\Omega_{\mathrm{Eff}}$. 

\begin{table}
\caption{\label{RabiTable} Experimental two-photon coupling strengths $\Omega_{mn}$ in units of $E_R/\hbar$. Some transitions are associated with two coupling strengths, which interfere with relative phase angle $\chi = 2\pi/3$; see text for details.}
\begin{ruledtabular}
\begin{tabular}{llll}
 & $X\rightarrow Y$ & $Y \rightarrow Z$ & $Z \rightarrow X$ \\
 \hline
$\mathbf{k}_1-\mathbf{k}_2$ & $0.61(10)e^{-i\chi}+ $ & $0.70(11)e^{-i\chi}+$  & $0.45(7)$ \\
~ & $0.47(7)$ & $0.50(8)$  &   \\
$\mathbf{k}_2-\mathbf{k}_3$ & $0.61(10)e^{i\chi}+$ & $0.65(10)e^{i\chi}$+  & $0.43(7)$ \\
~ & $0.47(7)e^{-i\chi}$ & $0.46(7)e^{-i\chi}$  &   \\
$\mathbf{k}_3-\mathbf{k}_1$ & $0.63(10)+$ & $0.58(9)+$ & $0.42(6)$ \\
~ & $0.49(8)e^{i\chi}$ & $0.42(6)e^{i\chi}$  &   \\

\end{tabular}
\end{ruledtabular}
\end{table}

\subsection{Flux lattice phases.}

Towards achieving a full optical flux lattice, we consider the issue of phase in the context of Section \ref{RamanInterference}, and provide details on a specific implementation.

\subsubsection{Effect of Raman interference.}
To evaluate the impact of the Raman interference phases on the flux lattice scheme, we sum the phases around the plaquette drawn in main text Fig. \ref{SetupFigure}(b), in which an atom makes a complete cycle $X\!\!\rightarrow\!\! Y\!\!\rightarrow\!\! Z\!\! \rightarrow\!\! X$. We write the total Rabi strength of transition $mn$, where $mn = \{XY,YZ, ZX\}$, as $\Omega_{mn}(\chi_{mn})=\Omega_{mn,1}+e^{i\chi_{mn}}\Omega_{mn,2} = A_{mn} e^{i\Phi_{mn}}$, with

\begin{equation}\label{RamanPhaseEqn}
\begin{split}
    A_{mn}&=\sqrt{\Omega_{mn,1}^2+\Omega_{mn,2}^2 +2 \cos(\chi_{mn})\Omega_{mn,1}\Omega_{mn,2}}\\
    \tan(\Phi_{mn})&= \frac{\sin(\chi_{mn})\Omega_{mn,2}}{\Omega_{mn,1}+\cos(\chi_{mn})\Omega_{mn,2}}
\end{split}
\end{equation}

\noindent Adopting the coordinate system of Eqn. \ref{E1Field} and Eqn. \ref{E2Field}, we note that electric field of $\mathbf{k}_3$ is $\mathbf{E}_3 = E_{30}  (-\sin(\phi) \cos(\theta) \hat{x} - \cos(\phi)\cos(\theta) \hat{y} + \sin(\theta) \hat{z}) e^{ i\mathbf{k}_3\cdot \mathbf{r}}$. We then find the phases $\chi_{mn}$ to be

\begin{equation}
    \begin{split}
\chi_{X Y}&=-\phi-\pi/2,\\
\chi_{Y Z}&=\phi+\pi/2,\\
\chi_{Z X}&=2\pi
    \end{split}
\end{equation}

\noindent With no AOM phases set, an atom encircling the plaquette experiences a phase pickup $\exp(i(\Phi_{XY}+\Phi_{YZ}+\Phi_{ZX}))$. Using Eqn. \ref{RamanPhaseEqn}, we find 

\begin{equation}
\begin{split}
\Phi =&\Phi_{XY}+\Phi_{YZ}+\Phi_{ZX}  \\
=&\arctan\Big(\frac{-\cos(\phi)\Omega_{XY,1}}{\Omega_{XY,1}-\sin(\phi)\Omega_{XY,2}}\Big) + \\
& \arctan\Big(\frac{\cos(\phi)\Omega_{YZ,1}}{\Omega_{YZ,1}-\sin(\phi)\Omega_{YZ,2}}\Big) 
\end{split}
\end{equation}

\noindent For equal coupling strength ratios $\Omega_{XY,2}/\Omega_{XY,1}= \Omega_{YZ,2}/\Omega_{YZ,1}$, we see that $\Phi=0$.

\subsubsection{Flux lattice phase table}

To make contact with proposal \cite{Cooper2013}, which dictates a $2\pi/3$ gauge flux, we provide Table \ref{PhaseMappingTable}, which shows the Raman coupling phases that can be imprinted by our AOMs. Consider the plaquette labeled $\alpha$ in main text Fig. \ref{SetupFigure}(c). As the atom encircles it in the counter-clockwise direction, we can sum the net acquired phase by reading off from Table \ref{PhaseMappingTable}:

\begin{equation}\label{PhaseSum}
    \Phi_\alpha = \underbrace{\phi_2 - \phi_3}_{\mathbf{q}_2} + \underbrace{\phi_3'-\phi_1'}_{\mathbf{q}_3} + \underbrace{\phi_1''-\phi_2''}_{\mathbf{q}_1}
\end{equation}

\noindent Choosing $\phi_2 = -2\pi/3$, $\phi_3=2\pi/3$, and all other phases 0, we see that $\Phi_\alpha = 2\pi/3\; \rm{mod}\; 2\pi$. With this choice, it can be similarly confirmed that $\Phi_\alpha=\Phi_\beta=\Phi_\gamma$. 

\begin{table}
\caption{\label{PhaseMappingTable} AOM-programmable phases of the flux-lattice scheme matrix elements. Each row corresponds to a momentum kick $\mathbf{q}_i$.}
\begin{ruledtabular}
\begin{tabular}{llll}
& $X\rightarrow Y$ & $Y \rightarrow Z$ & $Z \rightarrow X$ \\
\hline
$\mathbf{q}_1 = \mathbf{k}_1-\mathbf{k}_2$ & $e^{i(\phi_1-\phi_2)}$ &  $e^{i(\phi_1''-\phi_2'')}$  &  $e^{i(\phi_1'-\phi_2')}$ \\
$\mathbf{q}_2 = \mathbf{k}_2-\mathbf{k}_3$ & $e^{i(\phi_2''-\phi_3'')}$ & $ e^{i(\phi_2'-\phi_3')}$  &  $e^{i(\phi_2-\phi_3)} $\\
$\mathbf{q}_3 = \mathbf{k}_3-\mathbf{k}_1$ & $e^{i(\phi_3'-\phi_1')}$ & $ e^{i(\phi_1-\phi_2)}$  &  $e^{i(\phi_1''-\phi_2'')}$ \\
\end{tabular}
\end{ruledtabular}
\end{table}

\subsection{Light shifts.}

To resolve lattice dynamics, we require Rabi coupling strengths $\Omega$ on the order of the temperature of the atoms, or $\Omega \sim c \kappa$, where $\hbar \kappa$ is the initial spread in atomic momentum. On the other hand, to enable a unique correspondence between pairs of Raman beams $\mathbf{k}_i,\mathbf{k}_j$ and spin states $X,Y,Z$ requires a nonuniformity $\epsilon$ in their energy spacing such that $\epsilon\gg \Omega$. In alkali-like atoms, such shifts could be easily provided by magnetic bias fields. In alkaline-earth-like elements, the ground states are relatively insensitive to magnetic field. The Zeeman shift coefficient of ${}^{87}\mathrm{Sr}$ is only $185 $ Hz/G and lacks any significant quadratic component~\cite{Boyd2007}, which could provide the nonuniformity. 

We instead turn to an ac Stark shift beam called $\omega_{\mathrm{Lift}}$, tuned to $434.829943(5)~\rm{THz}$, with a detuning $\Delta =2\pi\times 700$ MHz from the hyperfine resonance line ${}^1S_0(F=9/2) \rightarrow {}^3P_1(F=9/2)$. The beam's ex-situ measured waists are $(\omega_x,\omega_z) = (350, 330) \mu \rm{m}$, much larger than the in-situ cloud size of $\approx 30 \mathrm{\mu m}$. This laser (linewidth 100 kHz) is fiber-coupled at the experiment to improve pointing stability. Its polarization is linear and aligned with the bias magnetic field. The resulting tensor light shift coefficient $\alpha_t$ is proportional to the detuning from the three hyperfine levels in ${}^3P_1$ as~\cite{Steck2021}:
\begin{equation*}
\begin{split}
\alpha_t(F,\omega) = & \sum_{F_e} (-1)^{F+F_e} \sqrt{\frac{40F(2F+1)(2F-1)}{3(F+1)(2F+3)}} \\
& \times \begin{Bmatrix} 1 & 1 & 2 \\ F & F & F_e \end{Bmatrix} \frac{|\langle F || \mathbf{d} || F_e \rangle|^2}{2\hbar \Delta_{F_e}}
\end{split}
\end{equation*}

\noindent where the sum runs over the hyperfine levels $F_e=\{7/2,9/2,11/2\}$, $F=9/2$ is the ground-state total angular momentum, $\Delta_{F_e}$ is the detuning from the hyperfine resonance, the curly-braces indicate the Wigner-6j symbol, and $|\langle F || \mathbf{d} || F_e \rangle|$ is the hyperfine dipole matrix element given by   
\begin{equation*}
\begin{split}
| \langle F || \mathbf{d} || F_e \rangle  | = & |\langle J || \mathbf{d} || J_e \rangle | (-1)^{F_e+J+1+I} \\
& \times\sqrt{(2F_e+1)(2J+1)} \begin{Bmatrix} J & J_e & 1 \\ F_e & F & I \end{Bmatrix} 
\end{split}
\end{equation*}

\noindent with $J=0$, $J_e=1$, nuclear spin $I=9/2$, and $|\langle J || \mathbf{d} || J_e \rangle |^2 = \Gamma_{689} \frac{2J_e+1}{2J+1} \frac{3\epsilon_0 \hbar c^3}{\omega_0^3}$ is the fine-structure dipole matrix element at 689.4~nm. The tensor ac Stark shift $\Delta \nu$ on an atom in state $|F,m_F\rangle$ due to a beam with intensity $I_l$, polarized linearly along the quantization axis, is given by $h\Delta \nu=-\alpha_t(F,\omega)I_l ((3m_F^2-F(F+1))/F(2F-1)$. For an experimental power of 182 mW and the measured beam waists, we expect energy splittings $\Delta \nu_{XY}= 171$ kHz and $\Delta \nu_{YZ}= 130 $ kHz.

\begin{figure}[t]
\includegraphics[width=\columnwidth]{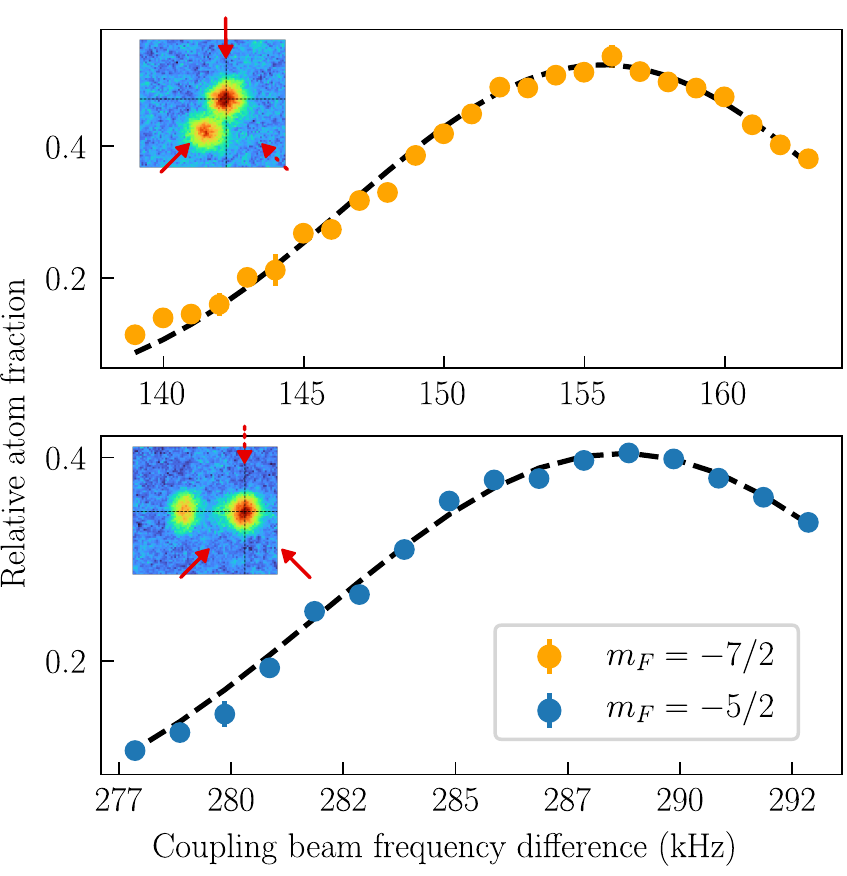}

\caption{Calibration of the ac Stark shift induced by the lift beam with a Fermi gas spin-polarized into $m_F=-9/2$. The beams were flashed on for $50 \mathrm{\mu s}$ ($75 \mathrm{\mu s})$ when transferring into the -7/2 (-5/2) state giving fit center 155.6 kHz (288.7 kHz); compensating for the recoil energy, this is 170.1 kHz (303.2 kHz). Insets show exemplary spin-momentum-resolved time-of-flight data, with the crosshairs indicating the location of $\mathbf{p}=0$. The solid red arrows indicate the involved beams, and the broken red arrows indicate a beam that is far-detuned from Raman resonance but providing its ac Stark shift.}
\label{EnergyLevelFigure}
\end{figure}

We measure the energy levels using Raman spectroscopy, as shown in Fig. \ref{EnergyLevelFigure}. Starting from a DFG polarized mostly into $m_F=-9/2$, we flash on the Raman beams $\mathbf{k}_1,\mathbf{k}_2,\mathbf{k}_3$ for varying times, with two beams probing a particular transition and the third beam detuned (but still providing its light shift). The frequency at which the population transfer is maximized determines the splitting~\cite{Han2019}. The data is fit with a Rabi-like model with temperature and pulse time as fixed parameters, varying center and Rabi frequency. After compensating for the recoil shift $4 E_R/h = 14.49 \mathrm{kHz}$, we find the energy differences $\delta \nu_{XY}=170.1 \pm 0.1\mathrm{kHz}$ and $\delta \nu_{ZX}=303.2 \pm 0.1 \mathrm{kHz}$, in excellent agreement with the expected light shift values. 

\emph{Lift beam Raman excitation.} We do not expect the lift beam to play a role in the dynamics by causing two-photon transitions. With the experimental lift beam intensity acting in conjunction with a 1.5 mW Raman beam, the lift beam's strongest two-photon Rabi frequency $\Omega_{\rm{Lift}} \approx 1$ MHz for the $m_F=-9/2$ (X) state, which we compute by summing the Rabi frequency contributions over all three $F_e$ hyperfine levels \cite{Kasevich1992}. The average two-photon detuning $\delta_{\rm{Lift}} = 1/2\times2\pi\times(-210+700)\; \rm{MHz}= 2\pi\times 245$ MHz, giving a $\pi$-time excitation probability   $\Omega_{\rm{Lift}}^2/(\Omega_{\rm{Lift}}^2+\delta^2_{\rm{Lift}}) \approx 10^{-5}$.

\subsection{Lifetime.}
\label{lifetime}

We model the lifetime in the spin-momentum lattice as the total rate of spontaneous emission. For a single-photon Rabi frequency $\Omega_{F_e}$ coupling a ground state to a hyperfine level ${}^3P_1(F=F_e)$ with detuning $\Delta_{F_e}$ and linewidth $\Gamma_{689}/2\pi = 7.4 \mathrm{kHz}$, the scattering rate is ~\cite{Steck2021}

\begin{equation*}
R_{\mathrm{sc}}=\sum\limits_{F_e}  \frac{\Omega_{F_e}^2/\Gamma_{689}}{1+(2\Delta_{F_e}/\Gamma_{689})^2+2 \Omega_{F_e}^2/\Gamma_{689}^2}
\end{equation*}

\begin{figure}[htp]
\includegraphics[width=0.9\columnwidth]{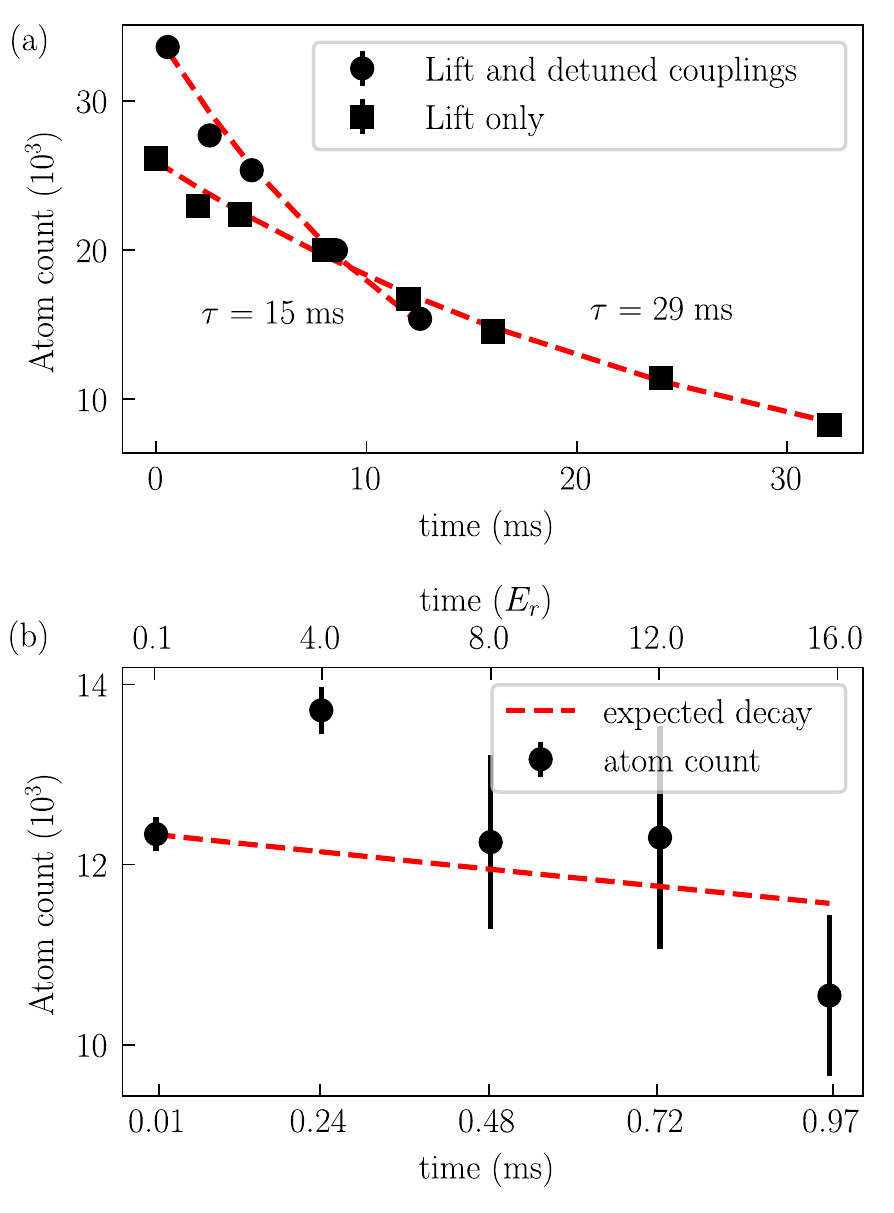}
\caption{ Lifetime measurements of the Fermi gas under different exposures to light. The lifetime in the ODT alone exceeds several seconds and is not shown. (a) (circles) Fermi gas lifetime with just the lift beam (see main text); (squares) both the lift beam and coupling beams at the experimental intensity, with the coupling beams detuned away from Raman resonance. (b) Total atom count measured during the experiment corresponding to the data in Figure 2 in the main text. From the measurement in (a), we plot the expected 15 ms decay curve. }
\label{LossFigure}
\end{figure}

Scattering arises principally from the lift and coupling beams. For the lift beam alone, we compute a characteristic timescale $\tau = 1/R_{sc} = 60~ \mathrm{ms}$ for the $m_F=-9/2$ (X) state, decreasing to $\tau = 23~\mathrm{ms}$ when also including the Raman coupling beams. In Fig. \ref{LossFigure}(a), we measure the impact each set of beams has on the overall lifetime. We find the lift beam imposes a $1/e$ decay time of 30 ms; the shorter than expected lifetime could be due to anti-trapping effects due to its positive scalar light shift, and could be compensated by a stronger ODT beam. Adding the Raman beams reduces the lifetime to 15 ms, much longer than the $\approx 1 \mathrm{ms}$ length of the current experiment. Fig. \ref{LossFigure}(b) compares the expected lifetime with the data in the Figure 2 of the main text. Since scattering scales as $R_{sc} \propto I/\Delta^2$ and Rabi frequency as $\Omega \propto I/\Delta$, we expect to improve the lifetime by further detuning the Raman beams from the current $\Delta/2\pi=210$ MHz. Changing to $\Delta/2\pi=400$ MHz would yield a 2-fold lifetime increase---reaching the limit imposed by the lift beam---with a commensurate increase in coupling power to maintain the same coupling strengths $\Omega_i$. Practically, this is achievable since the current total Raman power usage is only $\approx 4.5$ mW. Further improvements could be realized by choosing different lift beam detuning parameters, and could approach lifetimes of 70 ms \cite{Liang2021a}.

\begin{figure*}[t]
\includegraphics[width=\textwidth]{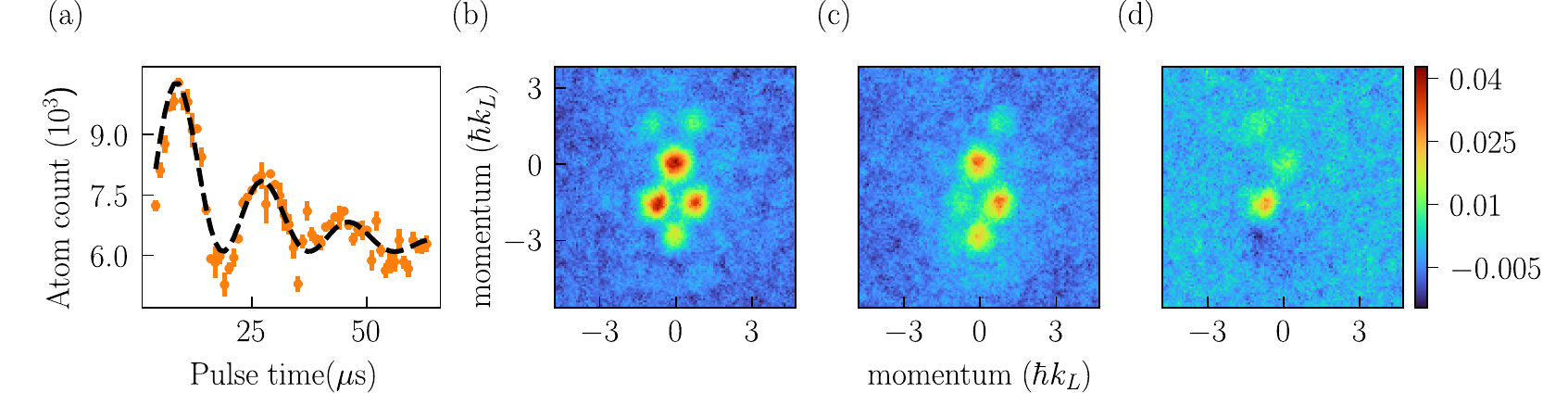}
\caption{ Spin-resolved imaging process. (a) Spin-blast Rabi frequency calibrated by electron shelving of the $m_F=-7/2$ (Y) state. The fitted single-photon Rabi frequency is $\Omega= 335(1)\; \mathrm{kHz}$. (b) Averaged atom shot of the SM lattice at sweep time $8E_R$. (c) Spin-blast beam resonant with $m_F=-7/2$ applied to the SM lattice. (d) Subtraction of images (b) and (c), removing the common-mode background and revealing the locations of $m_F=-7/2$ atoms. A negative-OD region forms at the locations of the blast-scattered atoms.  }
\label{SBlast}
\end{figure*}

\subsection{Spin-resolved imaging.}

Atoms released from the SM lattice naturally spatially resolve according to their spin and momentum; however, in cases where the starting state is not pure or for verification purposes, it is necessary to have a means to resolve the spins. Directly imaging the spins using narrow-line absorption imaging \cite{Stellmer2011} is impractical due to the low scattering cross-section and relatively low atom density of Fermi gases. Furthermore, since the atoms are moving after they are released from the SM lattice, the use of OSG separation is not desirable because it will (necessarily) disturb the momentum distribution. 

We instead visualize the spin dynamics with blast pulses \cite{Song2016}, propagating in the $xy$ plane, resonant with the narrow ${}^1S_0(F=9/2,m_F) \rightarrow {}^3P_1(F=9/2,m_F)$ transition at 689 nm. To be effective, the blast beam strength needs to be similar to the worst-case atomic Doppler shifts in the SM lattice $\Omega_{\mathrm{Doppler}} \approx  n \times 2\pi\times 4E_R = n \times 91$ kHz, where $n$ is the number of photons absorbed in the SM lattice. Simultaneously, to avoid exciting neighboring $m_F$ states, the blast strength must be smaller than the typical Zeeman splitting of the upper ${}^3P_1 (F=9/2)$ states, here $2\pi \times 790$ kHz. 

The beams are calibrated as in Fig. \ref{SBlast}(a) with electron-shelving \cite{Aman2019}: atoms are excited into ${}^3P_1$ with a 689 nm pulse for a time $t$, after which we apply a $4\; \mu \mathrm{s}$ pulse of 461 nm MOT light to blow away all remaining ground-state atoms. Atoms so ``shelved" in the ${}^3P_1$ state do not interact with the 461 nm light; instead, the fraction projected onto the ground state is measured with absorption imaging after a brief 2 ms time of flight. The overall decay curve matches the ${}^3P_1$ natural lifetime of $22\;\mu\mathrm{s}$. The SM lattice in Fig. \ref{SBlast}(b) is subjected to this beam in Fig. \ref{SBlast}(c), yielding the spin-resolved picture in Fig. \ref{SBlast}(d) after subtracting the two images.

Since the blast process relies on spontaneous emission, an atom can scatter (at most) a few photons before decaying to an adjacent $m_F$ state, which is then transparent to the narrow-linewidth blast beam. This means the momentum acquired by the targeted atoms is comparable to the momentum of the atoms in the SM lattice, so some overlap is inevitable---as the negative optical density region in Fig. \ref{SBlast}(d) shows. This effect can be particularly noticeable if the scattered atoms have strong geometric overlap with a region of interest, as seen in the $X$ row of main text Fig. 2, particularly the $8E_R$ panel, which entirely masks the cloud at $\mathbf{p}=-3\hbar k_L \hat{y}$.

Furthermore, although the beam strengths are chosen to be comparable to the Doppler shifts, the efficiency of these single-tone blast pulses still decreases with increasing SM lattice photon absorption. To overcome this, a larger bias magnetic field would enable the use of stronger blast pulses. To overcome the geometric overlap issue, tilting the blast pulses out-of-plane would push targeted atoms out of the imaging focus region, more effectively removing them from the images.

\end{document}